Concept: Collective Behaviour

BOX: The way in which an individual unit's actions is dominated by its neighbours so that all units simultaneously alter their behaviour to a common pattern.

## A question of scale

Tamás Vicsek

If you search for 'collective behaviour' with your web browser most of the texts popping up will be about group activities of humans, including riots, fashion and mass panic. Nevertheless, collective behaviour is also considered to be an important aspect of observed phenomena in atoms and molecules, for example, during spontaneous magnetization. In your web search, you might also find articles on collectively migrating bacteria, insects or birds; or phenomena where groups of organisms or non-living objects synchronize their signals or motion (think of fireflies flashing in unison or people clapping in phase during rhythmic applause).

What is the common factor, if any, in these seemingly diverse phenomena? The answer is that they happen in systems consisting of many similar units interacting in a relatively well-defined manner. These interactions can be simple (attraction/repulsion) or more complex (combinations of simple interactions) and can occur between neighbours in space or on an underlying network. Under some conditions, in such systems various kinds of transitions occur during which the objects (particles, organisms or even robots) adopt a pattern of behaviour almost completely determined by the collective effects of the other objects in the system. In a glass of water such a transition is the freezing of the liquid (the water molecules all order themselves into a well-defined crystalline structure). In a group of feeding pigeons randomly oriented on the ground, it is ordering themselves into a uniform flock while flying away after a big disturbance. Thus, the main features of collective behaviour are first, that an individual unit's action is dominated by the influence of its neighbours (the unit behaves differently from the way it would behave on its own); and second, such systems exhibit interesting ordering phenomena as the units simultaneously change their behaviour to a common pattern

The most important point I would like to make is that collective behaviour is applicable to a great many processes in nature, so is an extremely useful concept in many contexts. Understanding a new phenomenon usually happens by relating it to a known one: a more complex system is understood by analysing its simpler variants. In the 1970s there was a breakthrough in statistical physics when a deep theoretical understanding of a general type of phase transitions was achieved due to the invention of the so-called renormalization group method. This theory showed that the main features of transitions are insensitive to the details of the interaction between the objects in a system, thus, as an extreme case, orientational forces between atoms can result in ordering phenomena similar to those observed in groups of much more complex units.

Consider, for example, many thousands of people standing on a square and trying 'blind' to turn in the same direction. A nice example of collective behaviour would be if all of them faced the same way. Can they do it? Statistical physicists can predict for sure that they can't. They recall a theorem valid for particles with short-range ferromagnetic interactions stating that in two dimensions no long-range ordered phase can exist in such a system for any finite temperature and zero external field. So what happens in our example? Locally, people are looking almost in the same direction, but on a large scale, for example, seen from a helicopter — just like the little magnets — they locally form vortex-like directional patterns due to the small perturbations caused by human errors. Curiously, if the crowd is allowed to choose from a few discrete directions, the ordering can be realized. Surprisingly, another recent theory predicts that if the people are asked to move in the same direction they will be able to.

The approach of treating flocks, or even crowds, as systems of particles naturally leads to the idea of applying successful methods of statistical physics (such as computer simulations or theories on scaling) to the detailed description of the collective behaviour of organisms. Indeed, for the past couple of years — making use of the concept of collective behaviour — many interesting discoveries have been made for a rich variety of physical, animal and human behaviours. We now understand much better such collective phenomena as aggregation, swarming, network formation and synchronization. Quantitative prediction of the collective reaction of people to specific situations (panic, vehicular or pedestrian traffic (or elections?) should become feasible in the not too distant future.

Further reading